\documentclass[12pt]{article}

\title{\textbf{Maxwell's Demon and the Thermodynamics of Computation}}
\author{Jeffrey Bub\thanks{Email address: jbub@carnap.umd.edu} \\ 
\small \textit{Philosophy Department, University of Maryland, College 
Park, Maryland 20742, USA}}
\date{}

\normalsize
\begin{document}
\maketitle

\begin{abstract}
It is generally accepted, following Landauer and Bennett, that 
the process of measurement involves no minimum entropy cost, but the 
erasure of information in resetting the memory register of a computer 
to zero requires dissipating heat into the environment. This thesis 
has been challenged recently in a two-part article by Earman and Norton. 
I review some relevant 
observations in the thermodynamics of computation and argue that 
Earman and Norton are mistaken: there is in principle no entropy cost 
to the acquisition of information, but the destruction of information 
does involve an irreducible entropy cost.
\end{abstract}  

\bigskip

\section{Introduction}

Maxwell first introduced the demon in a letter to Tait (dated 
December 11, 1867; see \cite{Knott}, p. 214) and repeated the demon 
argument in his 1871 treatise, \textit{Theory 
of Heat}. He imagined a being capable of monitoring the positions and 
velocities of the individual 
molecules in a container of air at uniform temperature, 
divided into two chambers by a partition with a small aperture. 
While the mean velocity of the air molecules is uniform, the velocities of 
the individual molecules vary. The demon 
opens and closes the aperture so as to allow the faster 
molecules to move to one chamber and the slower molecules to the 
other chamber. In this way the 
temperature of one chamber is raised and the temperature of the other 
chamber lowered 
without any work being done, 
contradicting the second law of thermodynamics

As Maxwell indicated in an undated letter to Tait (see \cite{Leff}, 
p. 5, and \cite{Knott}), 
the point of the demon argument was to show that the second law `has 
only a statistical certainty.' The question then arises 
whether it is possible to design a perpetual motion 
machine that, over time, reliably exploits statistical fluctuations to 
convert heat from the environment into work. 

In 1912, 
Smoluchowski~\cite{Smoluchowski1, Smoluchowski2} showed that an 
automatic mechanism, like a spring-loaded trapdoor blocking an 
aperture between two chambers of a gas-filled container and capable of 
opening only one way, would be 
prevented by its own Brownian motion from functioning 
reliably as a one-way valve that allowed more energetic gas
molecules to accumulate in one chamber over time. The trapdoor would 
be heated by collisions with the gas molecules and open and close 
randomly, and these random fluctuations in the trapdoor motion would 
allow a molecule to pass from the hotter chamber to the 
colder chamber as often as a molecule pushes past the 
trapdoor from the colder chamber to the hotter chamber. So such a device 
could not function  as a perpetual 
motion machine capable of converting heat from the environment into 
work. In other words, a purely mechanical Maxwell's demon is 
impossible. 

In his seminal article~\cite{Szilard}, Szilard cites Smoluchowski 
(\cite{Smoluchowski2}, p. 89):
\begin{quote}
    As far as we know today, there is no automatic, permanently 
    effective perpetual motion machine, in spite of the molecuar 
    fluctuations, but such a device might, perhaps, function regularly 
    if it were appropriately operated by intelligent beings \ldots
\end{quote}
and states his objective as follows:
\begin{quote}
    The objective of the investigation is to find the conditions which 
    apparently allow the construction of a perpetual-motion machine of 
    the second kind, if one permits an intelligent being to intervene 
    in a thermodynamic system.
\end{quote}

The appropriate way to think about this question is to consider 
whether a mechanical demon incorporating a computer could work as a 
perpetual motion machine---that 
is, a device with information-gathering and information-processing 
abilities. Assuming the second law, 
the relevant question is: at what stage of 
the information-gathering or information-processing would the device 
fail?

The accepted position, before arguments by Landauer~\cite{Landauer} and 
Bennett~\cite{BennettTDcomp}, was that there is an irreducible entropy 
cost to measurement. The thesis that acquiring information involves a certain 
entropy cost, specifically at least $k\log 2$ per bit of information, 
was first proposed by 
Szilard~\cite{Szilard} and elaborated by Brillouin~\cite{Brillouin} 
in terms of a model in which the demon measures the positions of the 
gas molecules by shining light on them. 

Landauer and Bennett argued 
that measurement is in principle reversible and can be done 
without entropy cost. 
By contrast, they showed that there is an irreducible entropy cost to information 
destruction as opposed to information acquisition: resetting the 
memory register of a computer to zero involves an entropy cost. This result 
has important consequences for the thermodynamics of computation. 
Bennett~\cite{Bennettrevcomp}, Fredkin~\cite{Fredkin}, and 
Toffoli~\cite{Toffoli} showed that the most efficient 
computers, like Carnot engines, are reversible. It follows that the 
minimum 
energy required to carry out a computation does not depend on 
the complexity of the 
computation, but only on the number of bits of information in the 
output: if the output is 1 bit, one needs at least $kT\log 2$ of free 
energy to run the computation, which is used in resetting the memory 
to zero.

In a recent two-part article, Earman and 
Norton~\cite{EarmanNorton1,EarmanNorton2} reject the the use of 
information-theoretic notions to `exorcise' Maxwell's demon as misguided.
They argue that either the demon is a thermodynamic 
system governed by the second law, in which case no further 
assumptions about information and entropy are needed to save the 
second law from the demon, or the demon is not such a system, in 
which case no information-theoretic assumptions can save the 
second law. 

Earman and Norton call `Szilard's principle' the principle that 
acquiring information involves a minimum entropy cost; specifically, 
gaining information that distinguishes between $n$ equally likely 
states dissipates a minimum entropy of $k\log n$ into the environment. 
`Landauer's 
principle' is the principle that erasing this 
information from a memory register involves a 
minimum entropy cost of $k\log n$. They reject 
the prevailing view that the locus of entropy
dissipation required 
to compensate for the demon's entropy reduction is correctly identified by 
Landauer's principle not Szilard's principle, as associated with the erasure of 
information in the demon's memory not the acquisition of information. 
As they see it, both 
principles depend for their validity on the second law and are 
not incompatible. If the demon is a canonical thermal system, then 
either the process of measurement, or the process of erasing the 
demon's memory, or both, will involve entropy dissipation sufficient 
to prevent the demon from exploiting thermal fluctuations over time 
to convert heat from the environment into work. 

I shall argue that Earman and Norton are wrong: in principle, 
the process of measurement need not involve any entropy cost, but the erasure of 
information in the memory register of a computer cannot be achieved 
without a minimum entropy 
cost. In section 2, I briefly review some relevant observations in the 
thermodynamics of computation. In section 3, I discuss measurement, 
and in section 4, I show why resetting the memory register of 
computer to zero requires dissipating heat 
into the invironment.

\section{The Thermodynamics of Computation}

Here I review some relevant observations in the thermodynamics of 
computation, following the discussion in Feynman~\cite{Feynman}.

The fundamental principle in the thermodynamics of computation is 
that information should be conceived as physically embodied in the 
state of a physical system. So we can, for example, think of a 
message on a tape---a sequence of 
0's and 1's---as represented by a sequence of boxes, in each of which 
there is a 1-molecule gas, where 
the molecule can be either in the left half of the box (representing 
the state 0) or the right 
half of the box (representing the state 1). 

If we assume that the tape (the sequence of boxes) is immersed in a 
heat bath at constant temperature $T$, the amount of work, $W$, 
required to compress the 
gas in one of the boxes to half the original volume $V$ isothermally 
is:
\begin{eqnarray}
    W & = & \int_{V}^{V/2}pdV
    \nonumber \\
    & = & \int_{V}^{V/2} \frac{kT}{V}dV \nonumber \\
    & = & kT(\log (V/2) - \log V) \nonumber \\
    & = & -kT\log 2 
\end{eqnarray}
where $p$ is the pressure of the gas and $k$ is Boltzmann's constant.  
(Conventionally, work done by a gas in 
expanding is taken as positive. The 
negative sign here indicates that the work is done on the gas. 
Concepts such as temperature and pressure for a 1-molecule gas  
are understood in a time-averaged sense.) 

The total energy, $U$, of the gas is related to the free energy, $F$, and 
the entropy, $S$, by the equation:
\begin{equation}
    U = F + TS
\end{equation}
In an isothermal compression, the total energy of the gas remains 
constant, so:
\begin{equation}
    \Delta F = -T \Delta S
\end{equation}
This represents the heat energy dumped into the environment (the heat bath) 
by the work done 
during the isothermal compression. So the entropy of the 1-molecule gas 
changes in this thermodynamically reversible change of 
state by an amount:
\begin{equation}
    \Delta S = -k\log 2
\end{equation}
and the entropy of the environment is increased by 
$k\log 2$. Equivalently, there is a change of $kT\log 2$ in the 
free energy of the gas. 

In statistical mechanics, the entropy of a system in a certain 
thermodynamic state is introduced as a 
measure of the number of microstates 
available to the system in the thermodynamic state. Specifically, the 
entropy is taken as proportional to the logarithm of the number of available 
microstates, with the proportionality factor $k$. 
This contrasts with the analysis of thermodynamic 
quantities like temperature and pressure, which are defined as 
statistical averages over a 
distribution of molecular configurations. The sense in which the 
entropy of a thermodynamic system is an objective property of the 
system is nicely captured by Jaynes in the following 
statement (\cite{Jaynes}, quoted in \cite{Leff}, 
p. 17):
\begin{quote}
    The entropy of a thermodynamic system is a measure of the degree 
    of ignorance of a person \textit{whose sole knowledge about its
    microstate consists of the values of the macroscopic quantities 
    $X_{i}$ which define its thermodynamic state.} This is a 
    completely `objective' quantity, in the sense that it is a 
    function only of the $X_{i}$, and does not depend on anybody's 
    personality. There is then no reason why it cannot be measured in 
    a laboratory.
\end{quote} 

In terms of a statistical mechanical analysis, 
the kinetic energy of the 1-molecule gas is unchanged by the 
compression. The only change 
is that initially, before the compression, the molecule could be 
anywhere in the volume $V$, while after the compression the molecule is 
confined to the region $V/2$. Since the number of microstates available 
to the molecule in the volume $V$ at temperature $T$ is proportional to $V$, 
if the volume of the 
gas is halved at constant temperature, the number of available 
microstates is halved, because the molecule has access to only half the 
number of possible positions. So
the entropy is decreased by 
an amount equal to $k(\log V - \log V/2) = k\log 2$.

The information in a message can be 
defined as proportional to the amount of free energy required to reset 
the entire message tape to zero, in the sense that each cell 
of the tape---each 1-molecule gas in a box---is compressed to half its 
volume, reducing the number of available microstates by half. In 
appropriate units (taking logarithms to the base 2), it takes 1 bit of 
free energy to reset each cell to a zero value.

Clearly, if we already know whether the value of a cell is 0 or 1, 
there is no information contained in the cell. In terms of the above 
definition, if the value is 0, we do nothing to reset the cell; if 
the value is 1 so that the molecule is in the right half of the box, 
we can insert a partition trapping the molecule in the right half and 
then turn the box over. This involves no expenditure of free energy 
(assuming quasi-static, frictionless motion). So it is only if we do 
not know whether the molecule is in the left half of the box or the 
right half---if the specification of the thermodynamic state of the 
1-molecule gas is simply that the molecule is somewhere in the 
box---that free energy is 
required to trap the molecule in one half of the box. Evidently, it 
should make no difference whether the 
zero for the tape is defined as a sequence of 0's or a sequence of 
1's. But this will only be the case if the reset operation is 
understood as a compression, to be applied to a cell irrespective of 
the value of the cell, that is, as \textit{an operation applied in ignorance 
of the whether the molecule is in the left half or the right half}, 
after which we know where the molecule is.

\section{Measurement}

The essential feature of a measurement is that it establishes a correlation 
between the state of a system 
and the state of a memory register. Now, establishing a correlation 
between the states of the two systems is 
equivalent to a copying operation, and there is no entropy cost to 
copying. This can be seen as follows (\cite{Feynman}, p. 155):

Suppose we have two memory registers, for definiteness two tapes, 
$T_{1}$ and $T_{2}$, each considered as above to consist of a sequence 
of boxes containing one molecule, which can be either in the left half 
of the box ($L$), representing a 0, or the right half of the box ($R$), 
representing a 1. Suppose each tape is in the same state 
(the same sequence of 0's 
and 1's) and we would like to reset each tape to the zero state, 
which we take as a 
sequence of 0's. 

We can use the first tape to reset the second tape 
as follows: If the state of the first box in $T_{1}$ is 0, do 
nothing to the state of the first box in $T_{2}$. If the state of the 
first box in $T_{1}$ is 1, insert a partition trapping the molecule
in the right half of the first box of 
$T_{2}$ and invert the box. Continue in this way for the other boxes in the tape. 
We now have to reset the first tape. 

It follows that the entropy cost of the reset operation for two 
identical tapes
is the same as the cost for one tape. (There is no more information 
in a tape and a copy than in a single tape.) So the entropy cost of copying 
the first tape (seeing that these operations are reversible) must be 
zero. In principle, then, insofar as a 
measurement can simply be regarded as a copying operation, a 
measurement process need not involve any 
entropy cost, that is, it can be done without the 
expenditure of free energy. 

Of course, 
there are 
measurement procedures---procedures for establishing correlations 
between systems---that will involve dissipating entropy into the 
environment, such as the optical 
procedure considered by Brillouin. But there is no 
requirement in principle for a mechanical Maxwell's demon that 
incorporates an information processing device to use a light source to 
distinguish the molecules.

For a 1-molecule gas in a box, 
Bennett~\cite{BennettScAm} has proposed a mechanical 
measurement apparatus designed to 
determine which half of the box the molecule is trapped in 
without doing any work, hence with no entropy cost (assuming 
frictionless forces and quasi-static motion).  Earman and 
Norton (\cite{EarmanNorton2}, pp. 13--14) object that Bennett's 
apparatus would be subject the usual fluctuation 
phenomena, since it is a mechanical device governed by Hamiltonian 
mechanics and so must behave like a canonical thermal system. These 
fluctuations would prevent the device from functioning as a 
measuring instrument, for much the same reason that Smoluchowski's 
trapdoor would fail to function as a sorting device. 

Now, Bennett 
proposed his apparatus as an idealized 
reversible measuring device to illustrate 
the theoretical possibility of measuring and recording the position 
of a molecule without bouncing light off the molecule, 
and without involving any thermodynamically irreversible step. As a 
real apparatus, it would undoubtedly fail to work. But the argument 
that measurement does not have to be thermodynamically costly can be 
made without exhibiting a measuring instrument that does not dissipate
any heat into the environment. The essential point is simply that a 
measurement does nothing more than establish a correlation, and so is 
equivalent to a copying operation.

\section{Erasure}

Consider Bennett's entropy analysis of Szilard's 1-molecule 
engine in \cite{BennettScAm}. 
The apparatus consists of a box containing one 
molecule, with a movable piston at the left end and the right end. 
The box is in contact with a heat reservoir, so that the 1-molecule 
gas can expand isothermally against the pistons.
The demon can insert a partition that separates the box into two 
equal parts, left ($L$) and right ($R$). Initially, the demon's memory register 
is in a neutral or 
ready state, 0. The demon first inserts the partition and then measures 
the location of the molecule, whether it is in $L$ or $R$. 

The phase space of the 1-molecule gas can be partitioned into two equal 
regions, $L$ 
and $R$, and the phase space of the demon's 
memory register can be partitioned into three equal regions, 
corresponding to either $0$, or registering $L$, or 
registering $R$. This yields a partition into six equal regions for 
the phase space of the combined system: 
$(L,L)$, $(L,0)$, $(L,R)$, $(R,L)$, 
$(R,0)$, $(R,R)$, where the first element in each pair represents the 
state of the 1-molecule gas, and the second element represents the 
state of the memory register.

Initially, the molecule can be anywhere in the box and the memory 
register is set to 0, so the entropy of the combined system 
is $\log V$ (in appropriate units), where $V$ here is the 
volume of the phase 
space region $(L,0)\cup (R,0)$. We assume that the insertion and 
removal of the partition does not involve friction and can be done 
without any work. After the measurement (considered as a 
copying operation that involves no entropy cost), the molecule 
can be either in $L$ (in which case the memory registers $L$) or in 
$R$ (in which case the memory registers $R$), so the entropy of the 
combined system is:
\[
\log[(L,L) \cup (R,R)] = \log V
\] 

The demon now pushes the 
piston on the side that does not contain the molecule towards the 
movable partition, and removes the partition when the piston reaches 
it. This compression phase does not involve any work, since the 
piston is pushed against nothing and we are assuming no friction. 
The entropy of the combined system 
after the compression phase is still $\log V$. 

Next, the molecule 
pushes against the piston in an isothermal expansion phase, absorbing 
heat from the environment through the walls as it expands at constant 
temperature until the piston is pushed back to its original position. 
After the expansion, the molecule occupies the entire region of the 
box, and the memory registers either $L$ or $R$, so entropy of the 
combined system is:
\begin{eqnarray}
\log[(L,L) \cup (L,R) \cup (R,L) \cup (R,R)] & = & \log 2V
\nonumber \\
& = & \log V + 1
\nonumber
\end{eqnarray}
an 
entropy increase of 1 bit (taking base 2 logarithms). 
At the same time, the entropy of the 
environment is decreased by 1 bit. 

While the molecule in the 
box is now in its original state, somewhere in $L \cup R$, 
the memory register is not at $0$ but still registers $L$ or $R$. To 
reset the memory to $0$ requires compressing the phase space of the 
memory register. We might think of the memory register too as a 
1-molecule gas, partitioned into three regions, $L$, $0$, and $R$. To 
erase the information in this system, the pointer-molecule, which is 
in the region $L \cup R$, must be compressed to the region $0$. After this 
erasure or compression of the memory phase space, the entropy of the 
combined system is:
\[
\log[(L,0) \cup (R,0)] = \log V
\] 
\textit{By the second law}, this entropy decrease of 1 bit in the system must be 
accompanied by an entropy increase in the environment of 1 bit; that 
is, a minimum entropy of 1 bit must be dissipated to the environment 
in resetting the memory to $0$. 

Earman and Norton (\cite{EarmanNorton2}, pp. 16--17) argue that a 
computerized demon can be programmed to 
operate a 1-molecule Szilard engine without the need to erase 
information. They consider a memory register with two states instead 
of three, labelled 
$L$ and $R$. At the starting point of the program, the memory register 
is set to $L$. If the molecule is detected in the left half of the 
box, there is no change in the memory register. If the molecule is 
detected in the right half, the program switches the state $L$ to $R$. 
Then, depending on the state of the memory, one of two subroutines is 
executed. The $L$-subroutine implements the appropriate 
compression-expansion sequence (partition inserted, piston pushed in 
from the right, etc.) and ends by leaving the memory in the state $L$. 
The $R$-subroutine functions similarly but ends by resetting the memory 
to $L$. So after a complete cycle, the engine and demon are 
returned to the initial state with an entropy reduction of 1 bit, 
in violation of the second law. Neither subroutine involves erasure, 
because the resetting 
operation---$L$ to $L$ or $L$ to $R$---depends on the state of the 
memory register: it does not involve compressing the phase space of 
the register.

But this is precisely the point: under the constraints imposed by 
Earman and Norton, the demon has been reduced to an 
automatic mechanism analogous to Smoluchowski's spring-loaded trapdoor, and 
we already know that such a device cannot work. As 
Landauer (\cite{Landauer}; p. 189 in \cite{Leff}) remarks:
\begin{quote}
    This is not how a computer operates. In most instances, a 
    computer pushes information around in a manner that is independent 
    of the exact data which are being handled, and is only a function 
    of the physical circuit connections.
\end{quote}
The question at issue is at what stage of the information acquisition 
or information processing a computerized demon would fail as a 
perpetual motion machine, if we 
assume that the system is a canonical thermal system subject to the 
second law. The claim that information erasure, and not information 
acquisition or information processing, involves a 
minimum entropy cost depends on the observation that (i) measurement 
is essentially a copying operation with no entropy cost in principle, 
(ii) reversible computation is 
possible in principle, and 
(iii) erasure involves compressing the phase space of 
the physical system that functions as a memory register, which 
requires dumping heat into the environment.

A reset operation is logically irreversible, in the sense that the 
output does not uniquely determine the input (the mapping is 
many-to-one). If information is 
understood as physically embodied information, a logically 
irreversible operation 
must be implemented by a physically 
irreversible device, which dissipates heat into the environment. 
In \cite{Landauer}, 
Landauer considers a sequence of $n$ bits physically represented as an 
array of spins, initially all aligned in the positive $z$-direction, 
a state he designates as ONE. As the spins take up entropy from the 
environment, they become disoriented, so that each spin can be 
aligned either in the positive or in the
negative $z$-direction, with equal probability. Since the array can 
be in any one of $2^{n}$ states, the entropy can increase by $kn\log 
2$ (or $n$ bits, in appropriate units) 
as the initial information becomes thermalized. The reset operation 
RESTORE TO ONE is the opposite of thermalization: each bit is 
initially in one of two states and after the reset operation is in a 
definite state. Since the number of possible states for each bit has 
been reduced by half, the entropy is reduced by $k\log 2$ per bit. 
The entropy of a closed system, such as a computer with its own 
batteries, cannot decrease, so this entropy appears as heat dumped into 
the environment. 

Landauer(\cite{Landauer}; p. 192 in \cite{Leff}) remarks:
\begin{quote}
    Note that our argument here does not necessarily depend upon 
    connections, frequently made in other writings, between entropy 
    and information. We simply think of each bit as being located in a 
    physical system, with perhaps a great many degrees of freedom, in 
    addition to the relevant one. However, for each possible physical 
    state which will be interpreted as a ZERO, there is a very similar 
    possible physical state in which the physical system represents a 
    ONE. Hence a system which is in a ONE state has only half as many 
    physical states available to it as a system which can be in a ONE 
    or ZERO state.
\end{quote}

If all the bits in the array are initially in the ONE state, the 
reset operation RESTORE TO ONE involves no entropy change, and no 
heat dissipation, since no operation is necessary. Similarly, if all 
the bits are initially in the ZERO state, no entropy change is 
involved in resetting them all to the ONE state. (Recall Feynman's 
argument in section 1.) Landauer (\cite{Landauer}; pp. 192--193 in \cite{Leff}) 
notes that the reset operation would 
be different in these two cases: 
\begin{quote}
    Note, however, that the reset operation which sufficed when the 
    inputs were all ONE (doing nothing) will not suffice when the 
    inputs are all ZERO. When the initial states are ZERO, and we
    wish to go to ONE, this is analogous to a phase transition between 
    two phases in equilibrium, and can, presumably, be done 
    reversibly and without an entropy increase in the universe, but 
    only by a procedure specifically designed for that task. We thus 
    see that when the initial states do not have their fullest 
    possible diversity, the necessary entropy increase in the RESET 
    operation can be reduced, but only by taking advantage of our 
    knowledge about the inputs, and tailoring the reset operation 
    accordingly. 
\end{quote}

Earman and Norton(\cite{EarmanNorton2}, p. 16) cite these remarks by 
Landauer as justification for the claim that, 
in their program for a computerized demon with no erasure, 
neither subroutine 
involves erasure. Each subroutine is designed for a specific task: 
the $L$-subroutine ends by leaving the memory register in the state $L$, 
the $R$-subroutine ends by switching the state of the memory from $R$ 
to $L$. This is of course correct. But their example only succeeds in 
evading the issue: without a state-independent reset operation, their 
demon is reduced to an automatically functioning switching device, and 
the question raised by Szilard is not addressed.

\end{document}